# Nature of Planetary Matter and Magnetic Field Generation in the Solar System

by

**J. Marvin Herndon**
**Transdyne Corporation**
**San Diego, CA 92131 USA**

Communications: mherndon@san.rr.com    http://NuclearPlanet.com

**Abstract:** Understanding the nature of the matter comprising the Solar System is crucial for understanding the mechanism that generates Earth's geomagnetic field and the magnetic fields of other planets and satellites. The commonality in the Solar System of the matter like that of the inside of the Earth, together with common nuclear reactor operating conditions, forms the basis for generalizing the author's concept of nuclear georeactor geomagnetic field generation to planetary magnetic field generation by natural planetocentric nuclear fission reactors.

## Introduction

Currently active internally generated magnetic fields have been detected in six planets (Mercury, Earth, Jupiter, Saturn, Uranus, and Neptune) and in one satellite (Jupiter's moon Ganymede). Magnetized surface areas of Mars and the Moon indicate the former existence of internally generated magnetic fields in those bodies. Here evidence is presented attesting to the commonality of matter in the Solar System, which is like that of the deep-interior of Earth, and the suggestion is made that planetary magnetic fields generally arise from the same georeactor-type mechanism which Herndon [1] has suggested generates and powers the Earth's magnetic field.

There is clear evidence that certain planets contain internal energy sources. In 1969 astronomers discovered that Jupiter radiates into space more energy than it receives from the Sun. Verification followed, indicating that not only Jupiter, but Saturn and Neptune as well, each radiate approximately twice as much energy as they receive from the Sun [2, 3]. For two decades planetary scientists thought that they had considered and eliminated known planetary-scale energy sources, declaring "by default" or "by elimination" the observed internal energy must be a relic, leftover energy from planetary formation about 4.5 billion years ago [4, 5].

Applying Fermi's nuclear reactor theory [6], in 1992 Herndon [7] demonstrated the feasibility for planetocentric nuclear fission reactors as the internal energy sources for the gas-giant outer planets. Initially, he considered only hydrogen-moderated thermal neutron reactors, but soon thereafter demonstrated the feasibility for fast neutron breeder reactors as well, which admitted the possibility of planetocentric nuclear reactors in planets like Earth [8-10].

It is known that the Earth has an internal energy source at or near its center that powers the mechanism which generates and sustains the geomagnetic field. Applying Fermi's nuclear reactor theory [6], Herndon [8] demonstrated the feasibility of a planetocentric nuclear fission reactor as the energy source for the geomagnetic field. Subsequent state-of-the-art numerical simulations, made at Oak Ridge National Laboratory, verified his conjecture that the georeactor could indeed function over the lifetime of the Earth as a fast neutron breeder reactor and, significantly, would produce helium in the same range of isotopic compositions observed in oceanic basalts [11-13]. At this point, though, Herndon had only considered planetocentric nuclear fission reactors as planetary energy sources, not as mechanisms for generating planetary magnetic fields. Recently, he suggested that the georeactor is responsible, not only for powering the geomagnetic field, but for also being the mechanism responsible for actually generating the geomagnetic field [1].

The suggestion is herein made that the mechanism for generating planetary and satellite magnetic fields and for providing their requisite energy are one and the same, planetocentric nuclear fission reactors, like the Earth's georeactor [1]. That generalization is based upon fundamental considerations which demonstrate the commonality of highly-reduced, deep-Earth type matter, particularly within massive-cored planets of our Solar System, and the commonality of georeactor-like operating conditions.

## Nature of Planetary Matter

Only three processes, operant during the formation of the Solar System, are responsible for the diversity of matter in the Solar System and are directly responsible for planetary internal-structures, including planetocentric nuclear fission reactors, and for dynamical processes, including and especially, geodynamics. These processes are: (*i*) Low-pressure, low-temperature condensation from solar matter in the remote reaches of the Solar System or in the interstellar medium, which leads to oxygen-rich condensates; (*ii*) High-pressure, high-temperature condensation from solar matter associated with planetary-formation by raining out from the interiors of giant-gaseous protoplanets, which leads to oxygen-starved planetary interiors, and; (*iii*) Stripping of the primordial volatile components from the inner portion of the Solar System by super-intense solar wind associated with T-Tauri phase mass-ejections, presumably during the thermonuclear ignition of the Sun [14].

The constancy in isotopic compositions of most of the elements of the Earth, the Moon, and the meteorites indicates formation from primordial matter of common origin. Primordial elemental composition is yet evident to a great extent in the photosphere of the Sun and, for the less volatile, rock-forming elements, in chondrite meteorites. There is, however, a fundamental degree of complexity which has posed an impediment to understanding for more than half a century: Instead of just one type of chondrite there are three, with each type characterized by its own strikingly unique state of oxidation. Understanding the nature of the processes that yielded those three distinct types of matter from one common progenitor forms the basis for understanding much about planetary formation, their compositions, and the processes they manifest, including and especially magnetic field production.

Only five major elements [Fe, Mg, Si, O, and S] comprise at least 95% of the mass of each chondrite and, by implication, each of the terrestrial planets, and act as a buffer assemblage. Minor and trace elements are slaves to that buffer system and are insufficiently abundant to alter oxidation state. For decades, the abundances of major rock-forming elements ($E_i$) in chondrites have been expressed in the literature as ratios, usually relative to silicon ($E_i/Si$) and occasionally relative to magnesium ($E_i/Mg$). By expressing Fe-Mg-Si elemental abundances as molar (atom) ratios relative to iron ($E_i/Fe$), Herndon [15] discovered a fundamental relationship bearing on the genesis of chondrite matter, which has implications on the nature of planetary processes in our Solar System. The relationship obtained admits the possibility of ordinary chondrites having been derived from mixtures of two components, representative of the other two types of chondrite-matter. One component appears to be a relatively undifferentiated carbonaceous-chondrite-like primitive component, while the other, a partially differentiated enstatite-chondrite-like planetary component, appears to have originated from a large reservoir. Herndon suggested the partially-differentiated planetary component might be comprised of matter stripped from the protoplanet of Mercury, presumably by the T-Tauri solar wind during thermonuclear ignition of the Sun [15]. In other words, ordinary chondrite matter is not a primary building material for planets, although it might contribute a veneer to the terrestrial planets, especially to Mars.

Astronomical observations demonstrate conclusively that T-Tauri-type outbursts can be of sufficient magnitude to scour the terrestrial-planet region of our Solar System. Figure 1 shows a Hubble Space Telescope image of just such an outburst of the binary XZ-Tauri, taken in 2000. The white crescent embedded in the plume marks the leading edge of that outburst five years before. In five years the leading edge of the plume progressed a distance equivalent to 130 times the distance from our Sun to Earth.

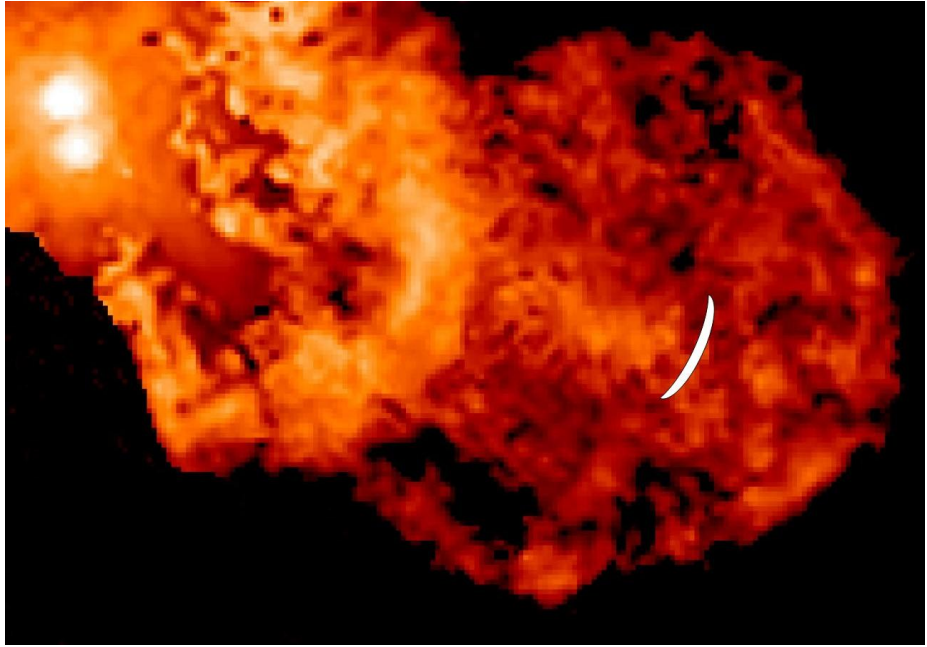

**Figure 1.** Hubble Space Telescope image of an outburst from the binary XZ-Tauri made in 2000. The white crescent shows the position of the leading edge of that plume in 1995, indicating a leading-edge advance of 130 A.U. in five years.

Much confusion has arisen from decades of making computational models based upon the erroneous assumptions that the mineral assemblage characteristic of ordinary chondrites formed in equilibrium in an atmosphere of solar composition at very low pressures, ca. $10^{-4}$ to $10^{-5}$ bars, and that ordinary-chondrite-like matter comprises planetary interiors.

Herndon and Suess [16] have shown that ordinary chondrite formation necessitates, not an atmosphere of solar composition, but instead an atmosphere depleted in hydrogen by a factor of about 1000. Subsequently, Herndon [17] showed the impossibility of ordinary chondrite-matter being in equilibrium with a gas of solar composition, and showed as well the necessity of some oxygen depletion relative to solar matter. Moreover, the ordinary chondrites appear, not primary, but rather as a secondary mixture, leaving only two types of primary matter, the oxygen-rich carbonaceous chondrite-type matter and the oxygen-starved enstatite chondrite-type matter [15].

As early as 1940, scientists, including the renowned Harvard geophysicist Francis Birch, built geophysics upon the premise that the Earth is like an ordinary chondrite, one of the most common types of meteorites observed impacting Earth, while totally ignoring another, albeit less abundant type, called enstatite chondrites. As Herndon [18] discovered in 1980, if the Earth is indeed like a chondrite meteorite as widely believed for good reasons, Earth is like an enstatite chondrite, not an ordinary chondrite. Imagine melting a chondrite in a gravitational field. At elevated temperatures, the iron metal and iron sulfide components will alloy together, forming a dense liquid that will settle beneath the silicates like steel on a steel-hearth. The Earth is like a spherical steel-hearth with a fluid iron-alloy core surrounded by a silicate mantle.

The Earth's core comprises about 32.5% of the planets mass. Only the enstatite chondrites, not the ordinary chondrites, have the sufficiently high proportion of iron-alloy that is observed for the core of the Earth, as shown in Figure 2. Moreover, components of the interior of the Earth can be identified with corresponding components of an enstatite chondrite meteorite: (1) The inner core being nickel silicide; (2) Earth-core precipitates CaS and MgS at the core-mantle boundary; (3) The lower mantle consisting of essentially FeO-free $MgSiO_3$; and, (4) The boundary between the upper and lower mantle being a compositional boundary with the matter below that boundary, the endo-Earth, being like an enstatite chondrite [18-21]. Those discoveries and insights led to a fundamentally different view of Earth formation, dynamics, energy production, and energy transport process [14, 22, 23].

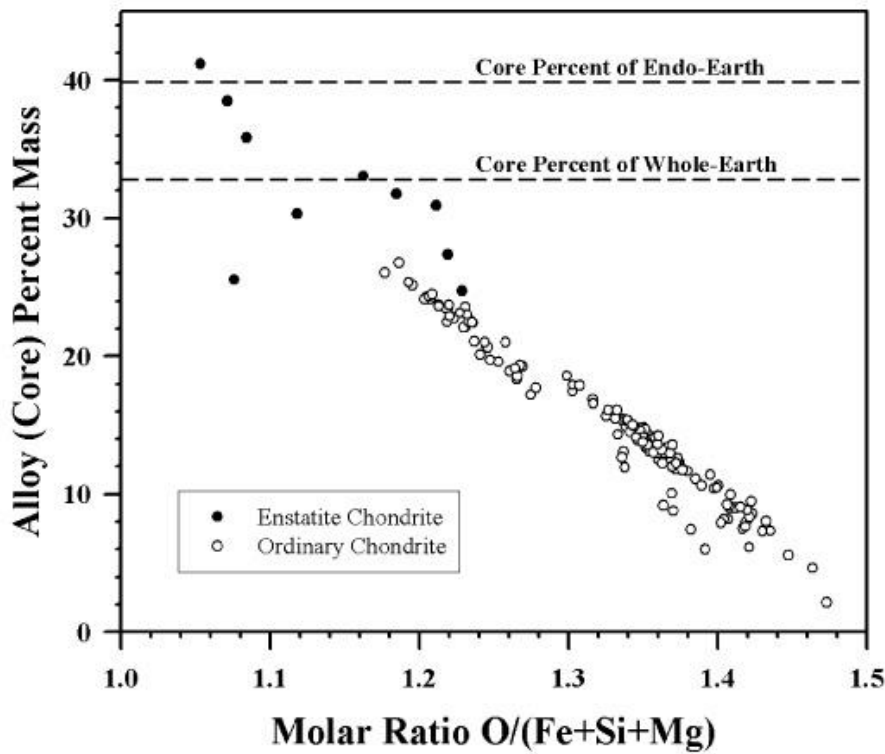

**Figure 2.** The percent alloy (mainly iron metal plus iron sulfide) of 157 ordinary chondrites and 9 enstatite chondrites plotted against a measure of oxygen content. The Earth as a whole, and especially the endo-Earth (core plus lower mantle) is like an enstatite chondrite and unlike an ordinary chondrite. For additional information, see [14, 21].

In the 1940s and 1950s, the idea was generally discussed about planets "raining out" from inside of giant gaseous protoplanets with hydrogen gas pressures on the order of $10^2$-$10^3$ bars [24-27]. But, in the early 1960s, scientists instead began thinking of primordial matter, not forming dense protoplanets, but rather spread out into a very low-density "solar nebula" with hydrogen gas pressures on the order of $10^{-4}$ to $10^{-5}$ bars. The idea of low-density planetary formation, often referred to as the "standard model", envisioned that dust would condense at fairly low temperatures, and then would gather into progressively larger grains, and become rocks, then planetesimals, and ultimately planets [28, 29].

These two ideas about planetary formation embody fundamentally different condensation processes which are the underlying cause for the two unique primary types of chondritic matter. The immediate implication is that both processes were operant during the formation of the Solar System. The relative extent and region of each process can be ascertained to some certitude from thermodynamic considerations together with planetary data. Even within present limitations, a consistent picture emerges that is quite unlike the so-called "standard model of solar system formation" [14].

From thermodynamic considerations it is possible to make some generalizations related to the condensation process in an atmosphere of solar composition. At the foundation, there are two dominant considerations, one essentially independent of pressure and one a strong function of pressure, which together are responsible for formation of the two primary types of Solar System matter.

In an atmosphere of solar composition, oxygen fugacity is dominated by the gas-phase reaction $H_2 + \frac{1}{2}O_2 = H_2O$ which is a function of temperature, but is essentially independent of pressure over a wide range of pressures where ideal gas behavior is approached. Oxygen fugacity controls the condensate state of oxidation at a particular temperature. At high temperatures the state of oxidation is extremely reducing, while at low temperatures it is quite oxidizing. The state of oxidation of the condensate ultimately becomes fixed at the temperature at which reaction with the gas phase ceases and/or equilibrium is frozen-in by the separation of gases from the condensate.

Condensation of an element or compound is expected to occur when its partial pressure in the gas becomes greater than its vapor pressure. Generally, at high pressures in solar matter, condensation is expected to commence at high temperatures, while at low pressures, such as $10^{-4}$ to $10^{-5}$ bar, condensation is expected to progress at relatively low temperatures at a fairly oxidizing range of oxygen fugacity. At low temperatures, all of the major elements in the condensate may be expected to be oxidized because of the great abundance of oxygen in solar matter relative to the other major condensable elements [30]. Beyond these generalizations, in this low-pressure regime, precise theoretical predictions of specific condensate compounds may be limited by kinetic nucleation dynamics and by gas-grain temperature differences arising because of the different mechanisms by which gases and condensate lose heat.

Among the thousands of known chondrites, only a few, like the famous Orgueil carbonaceous chondrite, have a state of oxidation and mineral components with characteristics similar to those expected as a condensate from solar matter at low pressures. Essentially all of the major elements in these few chondrites are oxidized, including sulfur.

The idea of planetary formation from a diffuse solar nebula, with hydrogen pressures on the order of $10^{-4}$ to $10^{-5}$ bar, envisioned that dust would condense at fairly low temperatures, and then would gather into progressively larger grains, and become rocks, then planetesimals, and ultimately planets. In the main, that picturesque idea leads to the contradiction of the terrestrial planets having insufficiently massive cores, because the condensate would be far too oxidized for a high proportion of iron metal to exist. But as evidenced by Orgueil and similar meteorites, such low-temperature, low-pressure condensation did in fact occur, perhaps only in the evolution of matter of the outer regions of the Solar System or in interstellar space, and thus may contribute to terrestrial planet formation only as a component of late-addition planetary veneer.

On the basis of thermodynamic considerations, Eucken suggested in 1944 core-formation in the Earth as a volatility-controlled consequence of successive condensation from solar matter in the central region of a hot, gaseous protoplanet with molten iron metal first raining out at the center [24]. Except for a few investigations initiated in the 1950s and early 1960s [25, 26, 31, 32], that idea languished when interest was diverted to Cameron's low-pressure solar nebula models [33].

On the basis of thermodynamic considerations, Herndon and Suess [34] showed at the high-temperatures for condensation at high-pressures, solar matter is sufficiently reducing, *i.e.*, it has a sufficiently low oxygen fugacity, for the stability of some enstatite chondrite minerals as shown in Figure 3. However, formation of enstatite-chondrite-like condensate would necessitate thermodynamic equilibrium being frozen-in at near-formation temperatures. At present, there is no adequate published theoretical treatment of solar-matter condensation from near the triple-point. But from thermodynamic and metallurgical considerations, some generalizations can be made. At the high temperatures at which condensation is possible at high pressures, nearly everything reacts with everything else and nearly everything dissolves in everything else. At such pressures, molten iron, together with the elements that dissolve in it, is the most refractory condensate.

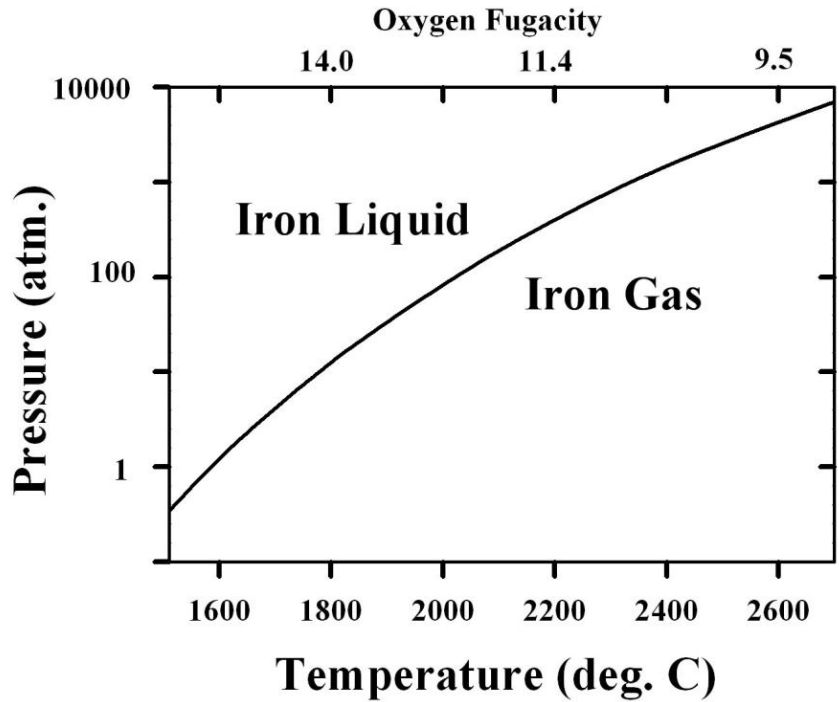

**Figure 3.** Phase boundary between iron gas and liquid iron metal as a function of temperature and pressure in an atmosphere of solar composition, calculated from thermodynamic data. Calculated oxygen fugacity values, independent of pressure, for selected temperatures are shown at the top.

Chondrite elemental abundances are nearly identical to solar element abundances for the relatively non-volatile rock-forming elements [30]. If Earth is like a chondrite element, as widely believed, then adding to Earth's mass the corresponding proportion of gaseous elements and those elements which form highly volatile compounds, calculated from solar abundances, yields an estimate of the mass of protoplanetary-Earth being in the range 275-305$m_E$, not very different from the mass of Jupiter, 318$m_E$. The formation of early-phase close-in gas giants in our own planetary system is consistent with observations and implications of near-to-star giant gaseous planets in other planetary systems [35-37]. It is thus reasonable to expect that the giant planets possess interior rock-plus-alloy kernels of enstatite-chondritic-like matter as they each possess internally generated magnetic fields.

In the absence of evidence to the contrary, the observed enstatite-chondritic composition of the terrestrial planets, as indicated by their massive cores, permits the deduction that these planets formed by raining out from the central regions of hot, gaseous protoplanets [14]. With the possible exception of Mercury, the outer veneer of the terrestrial planets may contain other components derived from carbonaceous-chondrite-like matter and from ordinary-chondrite-like matter. Mars, for example, may have an extensive outer veneer, while for Earth, it is ≤18% by mass. Satellites may possess an internal kernel of enstatite-chondritic-matter. The particular importance of enstatite-chondritic-matter derives from the highly reduced state of oxidation during formation, which forced certain oxyphile elements, such as uranium, into the alloy portion, rather than into the silicate.

### Planetary Magnetic Fields Generated by Nuclear Fission Reactors

Generation of magnetic fields in planets and satellites has long been conjectured to take place by the same mechanism responsible for generating Earth's magnetic field, a convecting fluid iron-alloy core dynamo. The commonality in the Solar System of the matter like that of the inside of the Earth, suggested here, is the basis for generalizing the concept of geomagnetic field generation by georeactor nuclear fission to planetocentric nuclear fission magnetic field generation in other planets and satellites. To appreciate the implications to other planets, it is beneficial to understand the circumstances related to our own planet, about which there is much more detailed information.

Elsasser [38-40] and Bullard [41] first adapted Lamor's self-exciting solar-dynamo concept [42] to explain the generation of the Earth's magnetic field. That mechanism is based upon the idea that convective motions in the Earth's fluid, electrically conducting core interacting with Coriolis forces produced by Earth's rotation, cause the fluid core to act like a dynamo, essentially a magnetic amplifier. Although being the subject of extensive investigations over more than a half century [43], there are fundamental problems with that concept.

For decades, the interior of Earth was erroneously assumed to be like an ordinary chondrite meteorite which, as was known, would have been too highly oxidized for the occurrence of radioactive elements in the core, although there was much discussion of the possibility that $^{40}$K might reside in the core. Realizing that the existence of the geomagnetic field demanded the presence of an energy source within the core, geophysicists often assumed, without corroborating evidence, that Earth's inner core was made of iron metal and that the inner core was growing, thus producing heat from the crystallization of iron supposedly to power the geomagnetic field.

Herndon discovered that the interior of the Earth is like an enstatite chondrite [18-21]. As a consequence of oxidation state, uranium in enstatite chondrites occurs in the alloy portion which corresponds to the Earth's core [44]. That observation led Herndon to disclose the background, feasibility and evidence of a nuclear fission georeactor at the center of the Earth as the energy source for the geomagnetic field [8-12, 45]. From fundamental considerations, as discussed above, there are only two types of primary, planet-building matter in Solar System of which only one type, like the deep interior of Earth, is capable of forming massive planetary cores. The commonality of planets with massive cores in the Solar System is indicative of their bulk compositions being of enstatite-chondrite-like matter and is the basis for generalizing the concept of planetocentric nuclear fission reactors as planetocentric energy sources and as the mechanism for generating planetary magnetic fields.

The geomagnetic field has existed for at least 3.5 billion years, as known from magnetic studies of rocks [46]. The almost universal belief that the geomagnetic field originates as a consequence of convection in the Earth's fluid core has led to little thought having been given to the possibility that there might be fundamental errors in the underlying assumptions, especially the assumption that convection in the Earth's alloy core can be sustained over extended times.

Nobel Laureate Chandrasekhar, an expert on convection [47], described convection in the following way [48]: "The simplest example of thermally induced convection arises when a horizontal layer of fluid is heated from below and an adverse temperature gradient is maintained. The adjective 'adverse' is used to qualify the prevailing temperature gradient, since, on account of thermal expansion, the fluid at the bottom becomes lighter than the fluid at the top; and this is a top-heavy arrangement which is potentially unstable. Under these circumstances the fluid will try to redistribute itself to redress this weakness in its arrangement. This is how thermal convection originates: It represents the efforts of the fluid to restore to itself some degree of stability." Understanding the clarity of Chandrasekhar's explanation has led to two reasons that convection in the Earth's fluid core is physically impossible.

Recently, Herndon pointed out that the Earth's fluid core is wrapped in an insulating blanket, a rock shell, the mantle, that is about 2900 km thick, and which has a considerably lower heat capacity, lower thermal conductivity, and higher viscosity than the fluid core [1]. Heat brought to the top of the core cannot be efficiently removed, so maintaining a significant difference in temperature between top and bottom of the core for extended periods of time, a requisite condition for long-term convection, is not possible.

Furthermore, as noted here, because of the over-burden weight, the Earth's core is about 23% more dense at the bottom than at the top [49]. Thermal expansion at the bottom cannot overcome such a great difference in density, meaning the Earth's core cannot become top-heavy, and thus cannot engage in convection. The implication is quite clear: Either the geomagnetic field is generated by a process other than the convection-driven dynamo-mechanism, or there exists another fluid region within the deep-interior of Earth which can sustain convection for extended periods of time.

Herndon has described the substructure of the Earth's inner core (radius 1250 km) as having at its center the georeactor, an actinide sub-core (radius 4 km) surrounded by a fluid or slurry sub-shell (radius 6 km) composed of fission products and products of radioactive decay [10], shown for example in Figure 4. The georeactor dimensions were very conservative estimates, and may in reality be as much as several times greater. The whole georeactor assembly is expected to exist at the center of Earth in contact with, and surrounded by, the nickel silicide inner core.

Convection in the fission product sub-shell is expected to be a stable feature of georeactor-like planetocentric nuclear reactors where nuclear fission produced heat is supplied directly to the base of the fission-product sub-shell whose outer boundary is a major heat sink. In Earth's georeactor, the outer boundary of the fluid sub-shell maintains contact with the semi-metallic, nickel silicide inner core, which acts as a heat sink, a thermal ballast, with reasonably good thermal conductivity to transport excess heat to the fluid iron-sulfur core, another heat sink. This arrangement enables the sub-shell's fluid to restore to itself, and to maintain an adverse temperature gradient and an enduring degree of convection stability. That arrangement also assures nuclear reactor stability.

In the micro-gravity environment at the center of Earth, georeactor heat production that is too energetic would be expected to cause actinide sub-core disassembly, mixing actinide elements with neutron-absorbers of the sub-shell, quenching the nuclear fission chain reaction. But as actinide elements begin to settle out of the mix, the chain reaction would re-start, ultimately establishing a balance, an equilibrium between heat-production and actinide settling-out, a self-regulating control mechanism.

A similar arrangement would be expected for planetocentric nuclear fission reactors in general. There is some question, however, as to what observable differences, if any, might arise if the outer boundary of a fluid fission product sub-shell were in contact with a fluid planetary core in the case of a yet un-precipitated inner core.

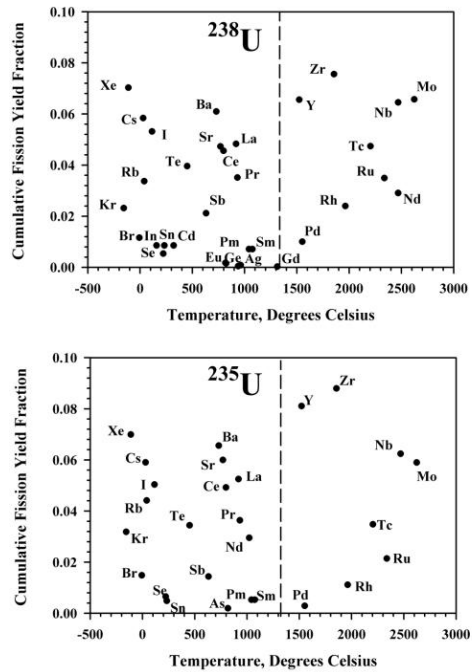

**Figure 4.** Cumulative yield fraction of $^{238}$U and $^{235}$U fast neutron fission product elements plotted vs. ambient pressure melting point of each respective element. The dashed vertical line represents the ambient pressure melting temperature of nickel silicide, $Ni_2Si$. Data from http://www.nucleonica.net. From [1].

The dynamo mechanism, thought to be responsible for generating the geomagnetic field, operates as a magnetic amplifier wherein, beginning with a small magnetic field, the combined motions of an electrically conducting fluid, driven by convection in a rotating system, amplify and maintain a more-or-less stable, much, much larger magnetic field. Herndon [1] has suggested for the Earth that geomagnetic field production occurs by the dynamo mechanism involving convection in the rotating, electrically conducting nuclear georeactor fission-product sub-shell, driven by nuclear fission energy produced in the georeactor sub-core. This fundamentally different concept is generally applicable to magnetic field generation in planets and their satellites, and appears to obviate the seeming paradox of Mercury's magnetic field.

Mercury is composed of enstatite-chondrite-like matter, as indicated by its massive core and by reflectance spectroscopic observations showing the regolith of Mercury to be virtually devoid of FeO, like the silicates of the enstatite chondrites [50]. Paradoxically, Mercury's small size would appear to preclude the existence of a fluid iron-alloy core, even if convection were possible, as the origin of its magnetic field. But, with Mercury's magnetic field generated by its planetocentric nuclear fission reactor, there is no paradox.

Seven of the eight Solar System planets either possess active, internally-generated magnetic fields or, in the case of Mars, display evidence of in the past having had such a magnetic field. Currently, The exception, Venus, has no active, internally-generated magnetic field for which there may be several; possible reasons: (1) The planet's rotation may be too slow for dynamo action; (2) The uranium nuclear reactor fuel may be exhausted; or, (3) The nuclear reactor may be temporarily shut down. Because the planet's surface is so hot, its rocks likely may never have been able to record evidence of an internally-generated magnetic field in its past as in the case of Mars. Venus' high surface temperature is thought by many to be the result of the greenhouse effect caused by its dense $CO_2$ atmosphere; the relative contribution from nuclear reactor heat is not possible to estimate due to various unknowns.

There are universal, inherent aspects to the generalized planetary dynamo concept involving convection in the rotating, electrically conducting the planetocentric nuclear fission-product sub-shell, driven by nuclear fission energy produced in the reactor sub-core. The power source and the magnetic field production mechanism are a single, self-contained unit that functions with the assurance of maintaining an adverse temperature gradient for sustained convection and nuclear reactor self-regulation. By virtue of its location, operating conditions are expected to be remarkably similar, *e.g.*, the microgravity environment, despite major differences in other aspects of the planets. The presence of a seed-field is assured through the radioactive $\beta^-$ decay of neutron-rich fission products and other ionizing radiation. The generality of magnetic field generation in planets and satellites by natural nuclear fission reactors is related to the generality of enstatite-chondrite-like matter as the primary planet-building material, as shown through the fundamental considerations presented here.